\documentclass[num-refs]{wiley-article}

\usepackage{siunitx}

\papertype{Original Article}

\title{Leveraging LLM-Respondents for Item Evaluation: a Psychometric Analysis}

\author[1]{Yunting Liu}
\author[2]{Shreya Bhandari}
\author[1]{Zachary A. Pardos}

\affil[1]{Berkeley School of Education, University of California, Berkeley}
\affil[2]{Electrical Engineering and Computer Science, University of California, Berkeley}

\corraddress{Berkeley School of Education, University of California, Berkeley}
\corremail{pardos@berkeley.edu}

\begin{document}

\begin{frontmatter}
\maketitle

\begin{abstract}
Effective educational measurement relies heavily on the curation of well-designed item pools (i.e., possessing the right psychometric properties). However, item calibration is time-consuming and costly, requiring a sufficient number of respondents for the response process. We explore using six different LLMs (GPT-3.5, GPT-4, Llama 2, Llama 3, Gemini-Pro, and Cohere Command R Plus) and various combinations of them using sampling methods to produce responses with psychometric properties similar to human answers. Results show that some LLMs have comparable or higher proficiency in College Algebra than college students. No single LLM mimics human respondents due to narrow proficiency distributions, but an ensemble of LLMs can better resemble college students' ability distribution. The item parameters calibrated by LLM-Respondents have high correlations (e.g. > 0.8 for GPT-3.5) compared to their human calibrated counterparts, and closely resemble the parameters of the human subset (e.g. 0.02 Spearman correlation difference). Several augmentation strategies are evaluated for their relative performance, with resampling methods proving most effective, enhancing the Spearman correlation from 0.89 (human only) to 0.93 (augmented human).
\end{abstract}
\end{frontmatter}

\textbf{Practitioner Notes}

\begin{itemize}
    \item What is already known about this topic
    \begin{itemize}
        \item Collection of human responses to candidate test items is common practice in educational measurement when designing an assessment
        \item Large Language Models (LLMS) have been found to rival human abilities in a variety of subject areas
        \item Data augmentation using AI has been an effective strategy for enhancing machine learning model performance
    \end{itemize}
    \item What this paper adds
    \begin{itemize}
        \item The first psychometric analysis of the ability distribution of a variety of open source and proprietary LLMs as compared to humans
        \item Finds that 50 LLM-Respondents produce item parameters similar to 50 undergraduate respondents
        \item Using LLM-Respondents to augment human response data gave mixed results
        
    \end{itemize}
    \item Implications for practice and/or policy
    \begin{itemize}
        \item The moderate performance of LLM-Respondents by themselves could provide a low-cost option for curating quality items for low stakes formative or summative assessments
        \item This methodology provides a scalable way to evaluate vast amounts of generative AI-produced items

    \end{itemize}
\end{itemize}

\section{Introduction}
Generating sets of well-functioning items for mathematics assessments often requires multiple iterations of calibration, involving extensive human participation. One of the most common techniques used in building an item pool is Item Response Theory (IRT), where ample response-level data is typically required for accurate item calibration and scaling \cite{konig2021benefits, kim2019application}. This process is time-consuming, costly, and significantly limits the rapid adaptation of educational assessments to different sets of students. For example, the PISA main survey requires between N = 250 to N = 750 respondents per item per country \cite{mazzeo2014linking}. Thus, the time and cost involved in obtaining responses from human respondents remains a significant area of resource expenditure.

With the advent of advanced AI technologies, novel ways to address these challenges have arisen. Recent developments in Large Language Models (LLMs) are achieving near-human performance \cite{katz2024gpt, chang2023survey, liu2023agentbench, achiam2023gpt}, leading to speculation about whether they can competently generate high-fidelity synthetic data without the traditional need for full data collection \cite{ding2024data, ye2024llm, whitehouse2023llm}. In our domain, we explore whether the capabilities of LLMs can be leveraged to provide responses resulting in psychometric properties similar to those derived from human respondents’ answers. This research is guided by three critical research questions:

\begin{itemize}
    \item \textbf{RQ1:} Which language model or configuration of language models best mimic human respondent abilities in mathematics, as measured by Item Response Theory (IRT)?
    \item \textbf{RQ2:} How do the psychometric properties of items fit to human responses compare to those fit to LLM-Respondents?
    \item \textbf{RQ3:} Can the augmentation of human respondent data with LLM-Respondent contributions yield item parameters comparable to those obtained from solely increasing human data collection?
    
\end{itemize}

If this approach is successful, it would mean that questions, including those produced via generative AI \cite{bhandari2023evaluating}, could be tested and evaluated en masse nearly instantly for use in a variety of educational contexts such as computer tutoring systems and other formative and summative assessment scenarios. 

In this study, we investigate the capabilities of various Large Language Models (LLMs), including GPT-3.5, GPT-4, Llama 2, Llama 3, Gemini-Pro, and Cohere Command R Plus, to generate assessment responses. Specifically, we prompt each model with 20 items sourced from the OpenStax Creative Commons textbook for College Algebra, producing 150 responses per model. Our analysis focuses on assessing whether these models can effectively replicate the response characteristics of our target population, which consists of undergraduate students in the United States, and on comparing the performance of these models against each other. To this end, we compare the model-generated responses to those obtained from U.S. undergraduates on the popular crowdsourcing platform, Prolific. 

\section{Related Work}

\subsection{Simulated Data in Educational Measurement and Educational Data Mining}
Analyzing examinee responses to test questions is indispensable in the field of measurement. While gathering real data can be time-consuming, costly, and often incomplete, simulation is a useful and economical technique since it can usually be done on a laptop without additional costs. Therefore, researchers commonly use simulation to validate models \cite{swaminathan200621}, compare different models \cite{harwell1996monte}, and evaluate estimation methods. In fact, among a random sample of publications in the field, 60\% of the studies used simulation, while the ratio for real data is just 41\% \cite{feinberg2016conducting}. Typically, a respondent distribution is specified, and then the item response level data (i.e., dichotomous or polytomous response) is simulated accordingly, with no thought process involved. Now, thanks to the advancement of generative AI, we can simulate some response level data and possibly gain insights into the cognitive structure behind the response process. While most work on simulation in educational settings is based on dialogue \cite{shaikh2024rehearsal, markel2023gpteach}, there are indeed some researchers conducting item response level data simulation. For example, Xu and Zhang \cite{xu2023leveraging} demonstrated the possibility of simulating student behavior based on assessment history. Lu and Wang \cite{lu2024generative} used insights from teachers to create generative students with various profiles, and then used the generative students' outcomes to guide item development. 

In the realm of educational data mining, gathering real learner data can pose privacy concerns \cite{hutt2023right} and present challenges with the costs of managing logged data \cite{jacob2015educational}. 
To address these challenges, researchers have at times leveraged simulated data. LLMs are often used to generate the training datasets needed to train and test other models. For example, by using pseudocode to generate synthetic datasets, researchers have been able to develop test cases of teaching activities to inform the development of a Teaching Outcome Model (TOM) \cite{ndukwe2018data}. Similarly, researchers have proposed using LLMs as data annotators to create synthetic data that can be used to train other models \cite{ding2024data}, mimicking the framework of Teacher-Student Learning (TSL) \cite{hu2022teacher}. Simulated data has also been used within the educational data mining community to evaluate latent trait models \cite{badrinath2021pybkt, beheshti2012methods}.

\subsection{Data Augmentation}
Data augmentation is a method often employed to increase the volume and diversity of data by generating new data from the existing set; it can also be applied to mitigate the 'incomplete data' problem \cite{seltzer1991use}. Having a limited number of data points often leads to weaker generalization capabilities, which can act as an obstacle to the effectiveness of studies \cite{kieser2023educational}. Thus, data augmentation is commonly used to enrich datasets and enhance their suitability for training models. For example, to augment image data, techniques such as resizing, rotating, and shifting images are frequently used \cite{xie2017data, shorten2019survey}. Researchers have also explored introducing noise in LLM training data \cite{xie2017data}, adding audio tracks or temporal shifts in speech recognition \cite{deng2000large, hannun2014deep}, and leveraging Generative Adversarial Networks (GANs) to generate training data for medical imaging \cite{frid2018gan}, ultimately creating datasets that are more generalizable and effective.

\subsection{OER and automation}
In recent years, the field of Open Educational Resources (OER) has seen significant growth and adoption \cite{henderson2018barriers}, allowing researchers to benefit from a corpus of educational resources at no cost and open materials that can be freely distributed, remixed, and adapted. With the rise of large language models (LLMs), the education sector is experimenting with automating the generation of these resources to reduce costs and enhance efficiency. In particular, there has been an emphasis on automatic item generation, hint generation, and skill tagging. For item generation, much research is focusing on utilizing the capabilities of LLMs to generate math questions either through template-based approaches \cite{shridhar2022automatic}, open-ended generation (Socratic style questions or math word problems) \cite{shridhar2022automatic, zhou2023learning, keller2021automatic, doi:10.1177/00472395231196532}, multiple-choice question generation \cite{9964056, nagasaka2020multiple, lee2024math}, or generation from structured formats (i.e., a bullet-point list) \cite{bhandari2023evaluating}. Hint generation has also been a focus, with researchers examining the effectiveness of LLMs in providing hints (i.e., worked solutions) to support learning in mathematics \cite{pardos2023learning, pardos2024chatgpt}, computer programming \cite{price2019comparison, rivers2014automating, piech2015autonomously, 10.1145/3231644.3231690, price2016generating, roy2016scale}, and various other STEM subjects. Additionally, studies have investigated human-AI collaboration in skill tagging, assessing its effectiveness across multiple languages and its speed and accuracy \cite{ren2024human, kwak2024bridging}. However, unlike these areas, the topic of using LLMs to simulate respondents remains under-researched. Thus, this paper aims to study the feasibility and effectiveness of LLMs in simulating respondents.

\section{Methods}

\subsection{Model Selection}
We selected six Large Language Models (LLMs) to generate responses that simulate answers from undergraduate college students in the U.S. to assessment questions. Our selection included GPT-3.5, GPT-4, Llama 2, the newer Llama 3, Gemini-Pro, and Cohere Command R Plus. These models were chosen for their varying levels of sophistication, reported accuracy on mathematics items, and widespread popularity, allowing us to simulate a broad spectrum of student abilities, from lower to higher academic proficiency \cite{touvron2023llama, plevris2023chatbots, team2023gemini}. For implementation, we utilized APIs for each model. As Llama does not offer a direct API, we accessed Llama 2 and Llama 3 via the Replicate API.

\subsection{Selection of Items and Prompt Engineering}
College Algebra was chosen as the subject because pre-authored questions were available under a CC BY license from an open textbook publisher, OpenStax\footnote{\url{https://openstax.org/details/books/college-algebra-2e}}. Additionally, we used a dataset from an earlier study that calibrated its item pool and had already collected responses from human participants via Prolific for 20 of the OpenStax College Algebra questions in Lesson 2.2: Linear Equations in One Variable \cite{bhandari2023evaluating}. This prior data collection contained some missingness in the data, so we were able to effectively use N >= 99 for all items.

 We distinctively formatted the questions, each prefixed with a label, in the following format: "Q1: <question1>" followed by a double newline, then "Q2: <question2>", and continued this pattern for all 20 questions. Specifically, the prompt was:

\begin{quote}
Q1: Given \( m=4 \), find the equation of the line in slope-intercept form passing through the point \( (2,5) \).

\bigskip
Q2: Find the slope of a line that passes through the points \( (2,-1) \) and \( (-5,3) \).

\bigskip
...

\bigskip
Q20: For the following exercises, solve the equation for x. State all x-values that are excluded from the solution set. \( 2-3/(x+4)=(x+2)/(x+4) \). Answer choices: Excluded values: -4 and x=-3; Excluded values: 4 and x=-3.
\end{quote}

This helped to ensure clarity and separation between items. We simulated 150 respondents for each LLM, as this was deemed the right sample size for conducting further analysis. To assess the accuracy of the responses, the second author of the study manually graded them and noted the accuracy for each one.

\subsection{Augmentation Procedure}
In a real-world case, sometimes only partial data is gathered. To explore the possibility of augmenting the data, we treated each human respondent in our dataset as a centroid, using only 50 human responses to represent the partial data. We then identified the nearest synthetic respondent from our pool of synthetically generated answers for each human centroid, allowing us to map the AI responses directly onto the characteristics of individual human responses. Next, we conducted a resampling procedure. First, we resampled a subset of 50 synthetic responses, selecting them based on the distribution of the models represented in the original matched 50, aiming to maintain the proportionality observed in this initial sampling. Lastly, we expanded this by resampling a subset of 100 synthetic responses using the same criteria.

\subsection{IRT analysis}

Contrary to sum-score analysis or percentage correct metrics, we plan to use Item Response Theory (IRT) to estimate the latent ability of human and LLM-Respondents. This method has several advantages over sum-score analysis \cite{demars2010item}. First, IRT assumes a latent trait theta, transforming all estimations onto a logit scale instead of the sum-score scale, which greatly improves measurement precision. Second, IRT provides person-level fit data, which can be done independently of other respondents. Lastly, and most relevant to our purpose, IRT maps both persons and items onto the same scales, enabling equating to be carried out without assuming population score distributions. In fact, IRT equating may be the best method when tests of differing difficulties are given to nonrandom groups of examinees who differ in ability \cite{cook1991irt}.

The simplest IRT model is often called the Rasch model or the one-parameter logistic model (1PL). The probability of individuals responding to a binary item (i.e., True/False) is determined by the individual's trait level and the difficulty of the item, which is often presented as:

$$P(X_{ij}=1|\theta_i,\beta_j) = \frac{\exp{(\theta_i-\beta_j)}}{1+\exp{(\theta_i-\beta_j)}}$$

where:

$X_{ij}$ refers to the response made by individual $i$ to item $j$. If the response is correct/true, then $X_{ij} = 1$.

$\theta_i$ refers to the trait level of individual $i$.

$\beta_j$ refers to the difficulty of item $j$.

\begin{figure}[ht!]
    \centering
    \includegraphics[width = 0.5\textwidth]{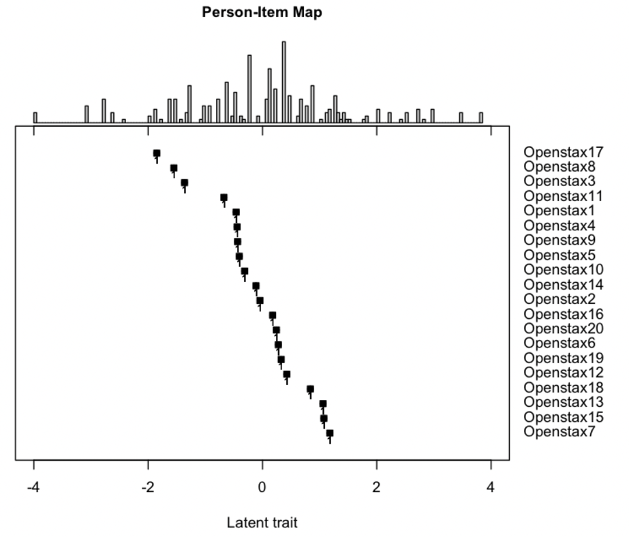}
    \caption{Item parameters calibrated by human respondents}
    \label{fig:wrightmap}
\end{figure}

A significant merit of using the Rasch model is that the estimates of latent trait and difficulty are mapped onto the same logit scale. The logits are interval units, which make the Wright Map (also known as the item-person map) a useful tool for presenting both item difficulties and person abilities arranged along the same logit scale. The location of item difficulty denotes the ability level at which the individual has approximately a 50\% probability of answering the item correctly. The results from human calibration using Rasch analysis are shown in Figure \ref{fig:wrightmap}. Items were ranked by their difficulties in ascending order from top to bottom \cite{wilson2023constructing}.

IRT analysis will be carried out for two purposes: Firstly, we will use the item parameters calibrated by human respondents as a standard so that the parameters for LLM-Respondents can be estimated. By treating item parameters as fixed, we will be able to compare the proficiency distribution of college students and LLM-Respondents. Within the fixed parameter calibration (FPC) realm, we will choose multiple weights updating and multiple EM cycles (MWU-MEM) as they are the most robust estimation methods \cite{kim2006comparative}. Secondly, separate calibrations will be carried out on human and AI respondent groups, informing us of the practicality of item calibration using AI generated data.

\section{Results}
\subsection{LLM-Respondent Simulation}
The initial item parameters for the item pool were calibrated on a group of current college students in the United States. Since multiple forms were used, there was missingness at random in the data, effectively resulting in N >= 99 for all items, satisfying the basic requirement for a Rasch analysis. The results are shown in Figure \ref{fig:wrightmap}. We then fixed the item parameters estimated from the model to obtain the proficiency estimates for the six AI models. We wondered whether the proficiency distribution of synthetic respondents is comparable to that of humans. Results show that most LLM proficiency distributions have a significant overlap with the human respondents. In particular, Llama 3 and GPT-3.5 have the highest mean proficiency distribution, which is higher than the human mean, indicating AI's greater proficiency in College Algebra compared to college students. The mean proficiency of GPT-4 is comparable to humans, while Cohere and Gemini are lower than humans. Llama 2 is the worst among all six models, suggesting its incapability in solving College Algebra problems.

\begin{figure}
    \centering
    \includegraphics[width = 0.5\textwidth]{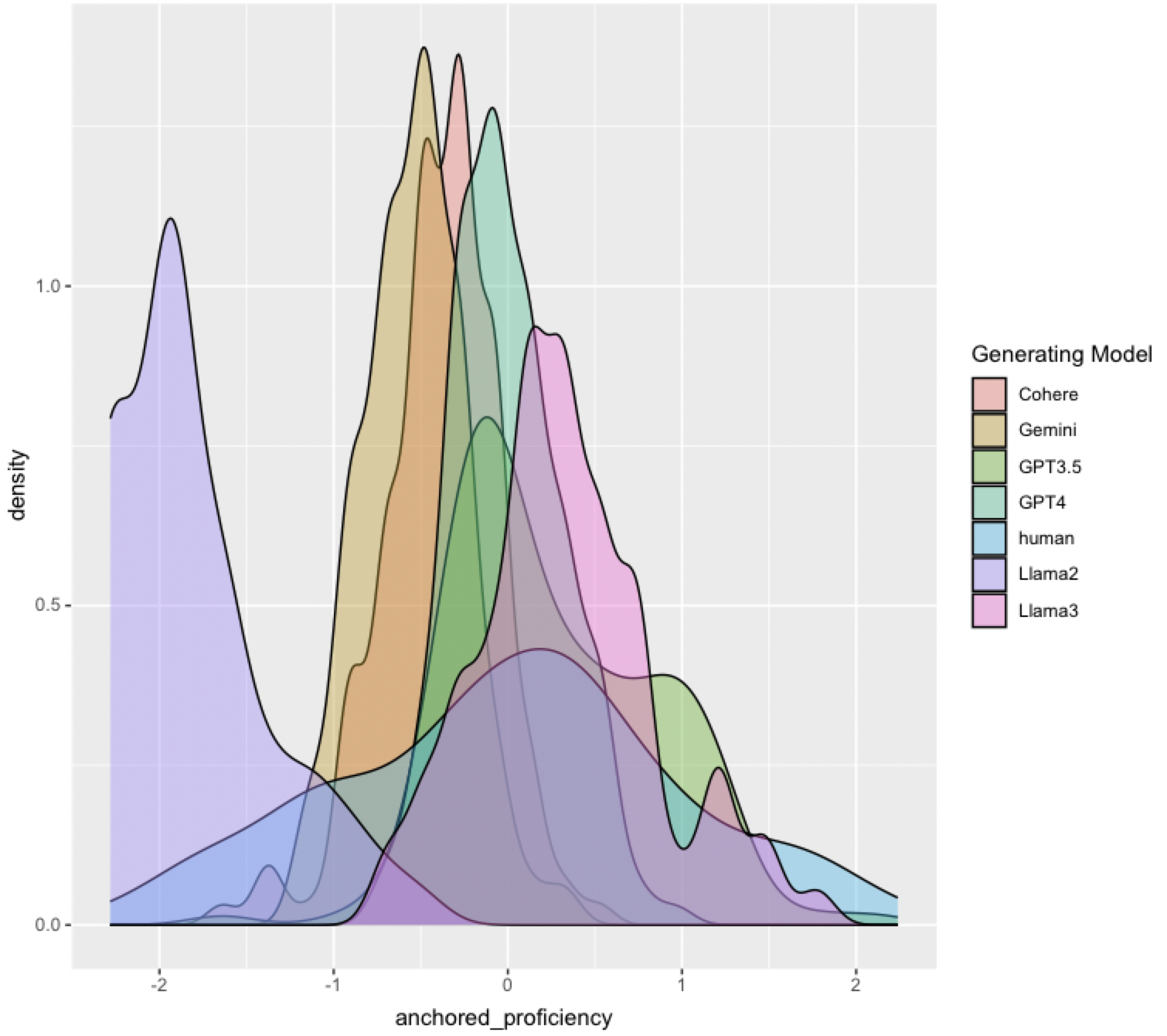}
    \caption{Proficiency distribution by Generating Models}
    \label{fig:proficiency_llms}
\end{figure}

Besides the findings on the average proficiency of each model, an interesting observation is that the variability in proficiency distribution for LLM-Respondents (standard deviation ranging from 0.29 to 0.58) was greatly reduced compared to the human distribution (SD = 0.98). Most AI distributions appear to be sharper than their human counterparts (kurtosis larger than 3). Detailed distributional information can be found in Table 1 and Figure \ref{fig:proficiency_llms}. Further analysis might explore ways to increase the variability in proficiency and response patterns.
\begin{table}[ht]
\centering
\caption{Statistical Measures of Response Distributions by Model}
\label{tab:response_stats}
\begin{tabular}{ccc}
\hline
\textbf{Generating Model} & \textbf{Mean} & \textbf{Standard Deviation (SD)}   \\    \hline
Cohere & -0.40 & 0.34 \\
GPT3.5 & 0.27  & 0.58 \\
GPT4   & 0.00  & 0.31 \\
Gemini & -0.54 & 0.29 \\
Llama2 & -1.81 & 0.44 \\
Llama3 & 0.37  & 0.51 \\
human  & 0.00  & 0.98       \\ \hline
\end{tabular}
\end{table}

We also performed item calibration on fully simulated data, comparing it against human-calibrated data and AI-calibrated data. Notably, the IRT estimated item difficulties show a significant correlation between AI-generated and human-generated data, especially for GPT-3.5 and GPT-4 (Pearson $\rho$ > 0.7), as shown in Table \ref{tab:response_stats}. Rank correlation (i.e., Spearman correlation) is also reported since not all distributions have the same mean, and the difficulties scale might not be an interval scale for different estimates. Given that one Gen-AI model might not have the power to represent the human distribution, we decided to use an ensemble method (also denoted as data augmentation) to create a representative pool of LLM-Respondents whose similarity with human respondents was boosted.

\begin{table}[ht]
\centering
\caption{Evaluation Metrics for Simulation and Augmentation experiments}
\label{tab:response_stats}
\begin{tabular}{cccc}
\hline
\textbf{Generating Source} & \textbf{Pearson $\rho$ } & \textbf{Spearman $\rho$ } & \multicolumn{1}{c}{\textbf{RMSE}} \\
\hline
GPT3.5     & 0.83              & 0.87             & 1.90                              \\
GPT4    & 0.72              & 0.78             & 2.27                             \\
Cohere    & 0.65              & 0.72             & 2.02                              \\
Gemini   & 0.68              & 0.61             & 2.67                              \\
Llama3    & 0.18              & 0.32             & 4.03                              \\
Llama2   & -0.01              &-0.07                 &     1.37                    \\
Experiment 1    & 0.92              & 0.89             & 0.55                              \\
Experiment 2    & 0.91              & 0.89             & 0.54                              \\
Experiment 3    & 0.93              & 0.93             & 0.71                              \\
Experiment 4    & 0.75              & 0.86             & 1.78              \\ \hline    
\end{tabular}
\end{table}
\subsection{Data Augmentation using LLM-Respondent}

Given that none of the LLM models have a proficiency distribution resembling that of humans, it is not feasible at this time to fully substitute human respondents with LLM-Respondents from a single LLM. However, LLMs could be used in a hybrid approach where half the respondents are human and half are LLM-Respondents. With this in mind, we propose three hybrid, or data augmentation, strategies listed below:

\begin{itemize}
\item An enlarged sample of 50 humans: 50 human respondents (examples) and 50 LLM-Respondents
\item A mixture of human respondents and resampled LLM-Respondents using proportions learned from humans
\item Fully LLM-Respondents using the mixing proportions learned from humans
\end{itemize}

The resampling analysis resulted in a set with significant variation in the proportions of each model used. GPT-3.5 was the most prevalent, comprising 36\% of the synthetic responses, followed by Llama 2 at 3\%. Gemini accounted for 12\%, while both Llama 3 and GPT-4 were represented at 8\% each. Cohere was the least represented model, constituting only 6\% of the responses.

To evaluate the relative performance of these three strategies, we proposed four experiments to test the effectiveness of different strategies on the item calibration process. The benchmark performance of the calibration is set by the human respondents; namely, we use all available data from human respondents to calibrate the item pool and gather item parameter estimates. Each experiment is designed to explore a different augmentation strategy. The proposed experiments are as follows:

\begin{enumerate}[label=Experiment \arabic*, labelsep=*, leftmargin=*]
    \item We use only half the number of the respondent pool and do the calibration (N = 50), representing a real-world scenario where there is a limitation in the budget, so only part of the intended respondents were collected.
    \item In addition to Experiment 1, we enlarge the data size by twice using augmentation strategies (N = 100).
    \item We use a mixture of human respondents and fully resampled data in a 1:1 ratio (N = 100).
    \item We use fully resampled data, with a size equal to the number of the benchmark dataset (N = 100).
\end{enumerate}

In terms of evaluation criteria, Pearson correlation and Spearman correlation with the benchmark condition are reported. We also use Root Mean Square Error (RMSE) to evaluate how accurate the new estimates are. RMSE measures the average difference between values predicted by a model and the actual values. $ RMSE=\sqrt{\frac{\sum_{i=1}^N\|y(i)-\hat{y}(i)\|^2}{N}}$. RMSE is always non-negative, and a value of 0 (almost never achieved in practice) would indicate a perfect recovery of the true/benchmark data.

\begin{figure}
    \centering
    \includegraphics[width = 0.5\textwidth]{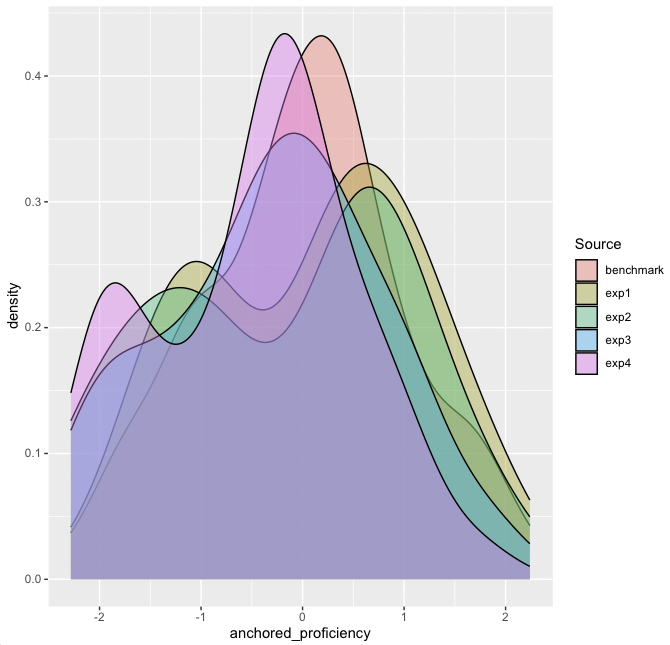}
    \caption{Proficiency distribution by Augmentation Experiments}
    \label{fig:proficiency_exps}
\end{figure}

Results show that when there aren't a sufficient number of respondents (Experiment 1), some difficulty estimates from the IRT analysis are not trustworthy, especially at the very ends of the respondent distribution (too-easy or too-hard items), resulting in a relatively high but not perfect correlation (Pearson $\rho$ = 0.92, Spearman $\rho$ = 0.89), and RMSE = 0.55. Among all augmentation strategies, Experiment 3 has the best result; it raised the Spearman $\rho$ from 0.89 to 0.93, indicating it is an effective strategy for recovering the order of item difficulties. However, since the mean center of the respondent distribution in Experiment 3 is tilted to the left ($\mu = -0.29$) compared to Experiment 1 ($\mu = 0.08$), the RMSE is larger as a universal shift is applied to all item parameters. Experiment 2 yields results comparable to Experiment 1, suggesting the strategy might not be beneficial in the current settings; practical reasons will be discussed in the Discussion section. As a validation, we also calibrated the respondent proficiency distribution using the fixed item parameter methods, and the results are shown in Figure \ref{fig:proficiency_exps}.

\section{Discussion and Conclusions}
In this study, we explored six different LLMs (GPT-3.5, GPT-4, Llama 2, Llama 3, Gemini-Pro, and Cohere Command R Plus) and various combinations using sampling methods to achieve psychometric properties similar to those from human respondents’ answers. Our findings are structured around three key conclusions: the proficiency of LLMs in approximating ability distributions (RQ1), item parameter correlation (RQ2), and the effectiveness of data augmentation (RQ3). Firstly, when comparing the proficiency between LLM responses and human responses, the results show that although some LLMs have comparable or even higher abilities in College Algebra, their distributions alone cannot fully represent the human distribution due to their narrow spread. Interestingly, this novel application of Item Response Theory (IRT) to LLM abilities reveals a first-of-its-kind distribution spread of abilities from multiple promptings, as opposed to the point estimates or percent-correct scores reported in other studies. Secondly, we compared the item parameters calibrated by AI responses and human responses, finding a relatively high Spearman correlation of 0.87 for GPT-3.5 and a lower 0.78 for GPT-4. Notably, GPT-3.5 emerged as the most human-like AI respondent, exhibiting Spearman correlations within 0.02 of those from 50 human respondents. Finally, since no single LLM currently has the capability to represent humans by itself, we explored ensembling approaches using three strategies. Among these, a mixture of human respondents and fully resampled data in a 1:1 ratio (Experiment 3) provided the best result, raising the Spearman correlation from 0.89 to 0.93. However, these augmentation results were mixed; while Spearman and Pearson correlations improved by 0.04 and 0.01, respectively, this approach substantially increased the RMSE. 

Our findings hold much promise for the automatic curation of items for tutoring systems. Simply put, it seems plausible to leverage AI respondents to curate an item pool that has a desirable spread of difficulty. Given the performance of 150 AI respondents from GPT-3.5 closely mirroring that of 50 humans (Experiment 1), AI respondents could be used as an initial filtering phase to reliably narrow down a larger item pool, and then have human respondents further refine the selection using the more manageable subset of items. This would significantly help optimize human resources, saving both time and money. For classroom environments, this research allows nimble testing of new questions, enabling selection of only quality assessments to present in the classroom.

\section{Limitations and Future Work}
Our study still has limitations. Firstly, we only utilized a single College Algebra lesson, which makes it difficult to generalize the results to other lessons within the same subject or to different subject areas. Additionally, our OpenStax item pool consists only of questions without images, figures, or tables because not all the LLMs support multimodal capabilities. Furthermore, in our experiment, the original human dataset displays bimodality. Therefore, when augmenting the data without resampling (Experiment 2), bimodality was also exhibited. While Experiment 3 mitigates this impact by using an effective resampling strategy, it would be beneficial to utilize a pool of human respondents that portray a normal distribution from the start. Due to leveraging prior data collection, we used the respondent sample size from their study, rather than calculating what the effective size should actually be.

In the future, analyses should be conducted to determine how valid the estimated proficiencies are when derived from a measurement tool calibrated by augmentation experiments versus those from benchmark data. To address this, we should investigate whether the structure of the tool with augmented respondents is the same as it is with the original human population \cite{ahmad2016assessing}. Typically, this is done by investigating the reliability of the measurement tool and conducting statistical analyses such as Explanatory Factor Analysis (EFA) or Confirmatory Factor Analysis (CFA) \cite{arafat2016cross}. 

The field should also aim to refine and expand methodologies to significantly improve the accuracy of responses generated by the models. In our study, we presented all 20 questions in a single prompt. However, it may be beneficial to question the model independently for each item. Although our initial trials with this technique for GPT-3.5 resulted in degenerate outcomes, this method could be extended to other models to assess its effectiveness more comprehensively. Additionally, more sophisticated prompt engineering techniques should be explored. Our prompt was simple and uniform across all models. However, it may be advantageous to customize prompts based on the model to better suit the specific strengths and design of each model. Incorporating few-shot learning by including one or two example responses in the prompt may also be effective. Finally, employing hallucination mitigation techniques such as self-consistency \cite{wang2022self}, in which the model is prompted multiple times for the same question and the modal response is selected, may help minimize erroneous outputs and thus increase accuracy. This technique has proven effective in reducing hallucination rates to near zero for answering mathematics questions \cite{pardos2024chatgpt}. Thus, expanding this technique, along with exploring other hallucination mitigation methods, may prove beneficial.

\end{document}